# Transformation of Amorphous Carbon Clusters to Fullerenes

*Alexander S. Sinitsa[1], Irina V. Lebedeva[2*], Andrey M. Popov[3], Andrey A. Knizhnik[4]*

[1] *National Research Centre "Kurchatov Institute", Kurchatov Square 1, Moscow 123182, Russia*
[2] *Nano-Bio Spectroscopy Group and ETSF, Universidad del País Vasco, CFM CSIC-UPV/EHU, San Sebastian 20018, Spain*
[3] *Institute for Spectroscopy of Russian Academy of Sciences, Fizicheskaya Street 5, Troitsk, Moscow 108840, Russia*
[4] *Kintech Lab Ltd., 3rd Khoroshevskaya Street 12, Moscow 123298, Russia*

**ABSTRACT.** Transformation of amorphous carbon clusters into fullerenes under high temperature is studied using molecular dynamics simulations at microsecond times. Based on the analysis of both structure and energy of the system, it is found that fullerene formation occurs in two stages. Firstly, fast transformation of the initial amorphous structure into a hollow $sp^2$ shell with a few chains attached occurs with a considerable decrease of the potential energy and the number of atoms belonging to chains and to the amorphous domain. Then, insertion of remaining carbon chains into the $sp^2$ network takes place at the same time with the fullerene shell formation. Two types of defects remaining after the formation of the fullerene shell are revealed: 7-membered rings and single one-coordinated atoms. One of the fullerene structures obtained contains no defects at all, which demonstrates that defect-free carbon cages can be occasionally formed from amorphous precursors directly without defect healing. No structural changes are observed after the fullerene formation, suggesting that defect healing is a slow process in comparison with the fullerene shell formation. The schemes of the revealed reactions of chain atoms insertion into the fullerene shell just before its completion are presented. The results of the performed simulations are summarized within the paradigm of fullerene formation due to self-organization of the carbon system.

## INTRODUCTION

Despite the fact that fullerenes have been known for almost 30 years, the detailed atomic mechanism of fullerene formation still remains a debated topic (see refs 1 and 2 for reviews). The most credible speculation is that the fullerene formation occurs as a spontaneous self-organization of the carbon system as the structural order arises through many local atomic structure changes induced by thermodynamic fluctuations without any certain thermodynamically stable intermediate structures. The detailed arguments which support this hypothesis are adduced in refs 1 and 2 along with the variety of initial systems for which fullerene formation is observed and stability of fullerenes in comparison with other carbon clusters. Namely, the diversity of these initial systems can be shown by the following examples: laser ablation of higher carbon oxides produces rings $C_{18}$, $C_{24}$, and $C_{30}$, which merge into large clusters and then transform to fullerenes[3,4]; transformation of bi- and tricyclic clusters consisting of 34-60 atoms to fullerenes has been observed in the drift tube[5,6,7,8]; fullerenes containing hundreds of atoms are formed by merging several $C_{60}$ fullerenes during ablation of the pure $C_{60}$ film.[9] Formation of fullerenes is possible not only at high temperature but also under electron irradiation. For instance, transformation of a graphene flake to a fullerene on a surface[10] and a nickel cluster surrounded by amorphous carbon to an endohedral metallofullerene[11] inside a carbon nanotube have been observed.

As for stability of fullerenes, calculations based on the density functional theory[12,13,14] and second-order Møller-Plesset perturbation method[12] show that the fullerene is the ground state for carbon systems which consist of more than 20 atoms. Theoretical considerations predict the upper limit of thermodynamic phase stability of fullerenes or multi-shell fullerenes (carbon onions) to be about 1000 atoms[15,16]. Thus fullerene formation should take place with a decrease of the potential energy regardless of the initial structure of the carbon system or the source of energy for bond rearrangement (heat treatment or electron or ion irradiation).

Since the experimental observation of individual bond rearrangement reactions for chaotic environment of fullerenes formation are hardly possible, the methods of atomistic modelling can be helpful. A set of works is devoted to molecular dynamics (MD) simulations of fullerene formation starting from carbon vapor[17-26], short carbon nanotubes[27,28], graphene flake[29,30] and a small nanodiamond cluster[31]. The simulation of processes similar to the fullerene formation such as the transformation of amorphous carbon[32,33] and nanodiamonds[32,34] to multishell carbon nanoparticles (onions) and graphitization of nanodiamonds of several nanometers in size[35,36] should also be mentioned. Here we consider yet another possible initial carbon structure which can be used for fullerene synthesis. Namely, the MD simulations performed on the basis of the reactive empirical potential show the transformation of an amorphous carbon cluster into a fullerene under heat treatment. Thus fullerene formation is possible as self-organization of any pure carbon system with an appropriate number of atoms. This fact supports the paradigm of fullerene formation as a result of self-organization.[1,2] The structural characteristics and energetics of the system during the transformation have been analyzed and have allowed to conclude that the transformation process takes place in two stages.

The paradigm of self-organization explains the high yield of abundant isomers of fullerenes (such as $C_{60}$ with icosahedral symmetry) by reactions which occur after the fullerene shell formation: emission of $C_2$ molecules[1,2,26,37-39], insertion of $C_2$ molecules[1,26,37-39] and, possibly, due to Stone-Wales (SW) reactions of bond rearrangement[1,40-44]. By now SW reactions[29,41,45,46] and reactions of insertion of $C_2$ molecules[39,47,48] that take place after the shell formation and finally lead to selection of abundant isomers have been carefully investigated. The SW reactions that can lead to defect healing have been also considered[29,44,49,50]. However little attention has been paid so far to the contribution of such reactions to the preceding process of shell formation. Up to now only collapse of large rings giving rise to 5-, 6- and 7-membered rings during the fullerene formation in carbon vapor[22,23,27,28] and insertion of five-coordinated atoms attached at the inner surface of the forming $sp^2$ shell in the transformation of a small nanodiamond cluster to a fullerene[31] have been discussed for the last stage of fullerene shell formation. The extensive studies of reactions at the last stage of shell formation are necessary to elucidate the detailed atomic mechanism of fullerene formation and thus to explain the high yield of abundant isomers of fullerenes. Here the schemes of insertion of atoms belonging to carbon chains attached at the outer surface of the forming $sp^2$ shell into the hollow $sp^2$ network that happens just before the end of the fullerene formation are presented. The types of defects remaining after the fullerene shell formation are also considered.

**COMPUTATIONAL DETAILS**

To study the transformation of amorphous carbon clusters into fullerenes under heat treatment we have performed reactive MD simulations using the first generation bond-order Brenner potential[51] (Brenner I) modified to reproduce graphene edge energies[52]. While a newer and more sophisticated second-generation Brenner potential[53] (Brenner II) is currently available, both of the original Brenner potentials are fitted to the experimental data on equilibrium distances, binding energies and stretching force constants of small hydrocarbons, graphite and diamond, which means they do not necessarily perform well for carbon nanostructures. With intention to provide the potential suitable for such condensed carbon phases, the training set of ReaxFF[54] force field, originally fitted to the quantum chemistry data on energetics and reactions of hydrocarbons[55], has been recently supplemented by the results of density functional theory (DFT) calculations on equations of state for diamond and graphite, formation energies of defects in graphene and heats of formation of amorphous carbon clusters[56]. To choose the adequate potential for our simulations we have tested these potentials with respect to the quantities relevant for the transformation of an amorphous carbon cluster into a fullerene. In such a process, the energy difference between the initial and final structures, i.e. the driving force, depends on the relation between the energy cost of undercoordinated (e.g., two-coordinated) carbon atoms compared to the three-coordinated ones and the penalty in the elastic energy coming from formation of the fullerene shell. Therefore, it is highly important that the interatomic potential employed is able to describe properly such characteristics as the elastic energy of fullerenes and graphene edge energies.

To compare the performance of different versions of the Brenner potential and the recent C-2013 version of ReaxFF we have calculated elastic energies of $C_{60}$ and $C_{70}$ fullerenes, i.e. the energy per atom relative to that in graphene, and graphene edge energies. Armchair, zigzag and reconstructed zigzag edges are considered, where the latter is obtained from the zigzag edge by transformation of all hexagon pairs at the edge into pairs of heptagons and pentagons[57–60]. The edge energies per unit edge length are found through calculations for wide graphene ribbons (of more than 4 nm width) with periodic boundary conditions along the ribbon axis as $E_{\text{edge}} = \left( E_{\text{ribbon}} - N_{\text{ribbon}} \varepsilon_{\text{gr}} \right) / L_{\text{edge}}$, where $E_{\text{ribbon}}$ is the ribbon energy, $N_{\text{ribbon}}$ is the number of atoms in the ribbon per unit cell, $\varepsilon_{\text{gr}}$ is the energy of bulk graphene per atom and $L_{\text{edge}}$ is the total edge length per unit cell. As seen from comparison with the results of spin-polarized DFT calculations in the local density approximation (LDA)[61] and in the generalized gradient approximation with the exchange-correlation functional of Perdew, Burke and Ernzerhof (PBE)[62] (see Table 1, a more detailed review on DFT calculations of graphene edge energies is available in Ref. 63), the first-generation Brenner potential describes the characteristics of interest much better than the second-generation Brenner potential, which fails even to describe the correct order of graphene edge energies. ReaxFF is rather accurate in the elastic energies of the fullerenes but also erroneously assigns the lowest energy to the unreconstructed zigzag edge. Setting the parameter $F(1,2,2) = -0.063$ in the first-generation Brenner potential (for the second set of parameters from paper[51]), as proposed in our previous paper[52], further improves the values of the graphene edge energies. Therefore, the modified first-generation Brenner potential able to describe the elastic energies and energies of undercoordinated atoms at the same level of accuracy seems to be the optimal choice for simulations of transformations of carbon clusters.

Table 1. Energies of the reconstructed zigzag ($E_{ZZ(57)}$), unreconstructed zigzag ($E_{ZZ}$) and armchair edges ($E_{AC}$) of graphene (in eV/Å) and elastic energies of $C_{60}$ ($E_{C60}$) and $C_{70}$ ($E_{C70}$) fullerenes (in eV per atom) obtained using different reactive interatomic potentials and by DFT calculations.

| Method | $E_{ZZ(57)}$ | $E_{AC}$ | $E_{ZZ}$ | $E_{C60}$ | $E_{C70}$ |
|---|---|---|---|---|---|
| Original Brenner I (Ref. 51) | 0.937 | 1.000 | 1.035 | 0.39 | 0.34 |
| Modified Brenner I (Ref. 52) | 1.147 | 1.247 | 1.494 | 0.39 | 0.34 |
| Brenner II (Ref. 53) | 1.124 | 1.091 | 1.041 | 0.55 | 0.49 |
| ReaxFF$_{C-2013}$ (Ref. 56) | 1.110 | 1.207 | 1.089 | 0.37 | 0.33 |
| DFT (LDA) | 1.147[a], 1.09[b] | 1.202[a], 1.10[b] | 1.391[a], 1.34[b] | | |
| DFT (PBE) | 0.98[c], 0.965[d] | 1.02[c], 1.008[d] | 1.15[c], 1.145[d] | 0.36[e] | 0.32[e] |

[a] Ref. 57; [b] Ref. 58; [c] Ref. 59; [d] Ref. 60; [e] Ref. 52

The adequacy of the modified first-generation Brenner potential used in the present paper has been already checked by simulations of the transformation of graphene flakes with nickel clusters attached into nickel heterofullerenes under electron irradiation[52]. The potential was also previously applied for modeling of carbon nanotube cutting catalyzed by nickel under electron irradiation showing the qualitative agreement of structure evolution with the experimental observations and correctly predicting the key stages of the cutting.[64]

MD simulations have been carried out using the in-house MD-kMC[65] (Molecular Dynamics – kinetic Monte Carlo) code. The periodic boundary conditions are applied to the simulation cell of 20 nm × 20 nm × 20 nm size with a single amorphous cluster. The procedure for generation of initial structures of amorphous carbon clusters is described in Supporting Information. The velocity Verlet integration algorithm[66,67] is used and the integration time step is 0.6 fs. The temperature is maintained at $T = 2500$ K by the Berendsen thermostat[68] with the relaxation time of 0.3 ps. Since the Berendsen thermostat can transfer the kinetic energy from the internal degrees of freedom to the global motion and rotation of the cluster[35], the linear and angular momenta of the system are set to zero every 0.3 ps.

Thermodynamic fluctuations have a considerable influence on non-equilibrium processes in nanoscale systems that can lead to different behavior for similar starting conditions. For example, MD simulations show that different metal-carbon nanoobjects form from a graphene flake with a metal cluster attached under electron irradiation[52] or from a metal cluster surrounded by amorphous carbon both under heating and electron irradiation[11] starting from the same initial structure. Only a minor part of MD simulation runs starting from carbon vapor at certain conditions leads to fullerene formation.[22,23] Thus at least several tens of simulation runs are necessary to obtain statistically significant results for non-equilibrium behavior of a nanoscale system. In total, we have performed 47 MD simulation runs of about 1 μs duration for slightly different initial structures of amorphous carbon clusters with the even number of atoms ranging from $C_{58}$ to $C_{74}$ and 25 MD simulation runs of about 400 ns duration for the same initial structure of amorphous carbon cluster $C_{66}$.

To describe the evolution of the system structure during the simulation runs we use the following definitions. We define a fullerene as a hollow closed carbon shell with the $sp^2$ structure without any carbon chains attached to the shell at one or both ends. Chains are defined as two or more connected two-coordinated and one-coordinated atoms. We assume that the fullerene shell may contain defects of the atomic structure such as single one-coordinated carbon atoms and rings not only of 5 or 6 atoms but also of 7 atoms (hereafter referred to as 5-rings, 6-rings and 7-rings, respectively). To identify chains and rings the topology of the carbon bond network is analyzed on the basis of the "shortest-path" algorithm.[69] Two carbon atoms are considered as bonded if the distance between them does not exceed 1.8 Å.

The following set of parameters is monitored in time to study the details of the transformation process: the potential energy of the system per atom, $E$, the numbers of 5-rings, $N_5$, 6-rings, $N_6$, 7-rings, $N_7$, and other rings (3-, 4-, 8- or 9-rings), $N_r$, the number of one-coordinated and two-coordinated atoms in chains, $N_c$, and the numbers of three-coordinated atoms which belong to the $sp^2$ network and to the amorphous domain, $N_3$ and $N_{3a}$, respectively. The latter two types of three-coordinated atoms are introduced to follow the build-up of the smooth and nearly spherical $sp^2$ shell from the disordered structure. To distinguish the $sp^2$ and amorphous domains the local normal to the shell around each three-coordinated atom (i.e. to the plane going through the three neighbors of the atom) is analyzed. While in the ordered $sp^2$ domain this local normal changes its orientation continuously from one atom to another, in the amorphous domain there is little correlation in the orientation of the local normal for the adjacent atoms. Therefore, to calculate the numbers $N_3$ and $N_{3a}$ each three-coordinated atom is assigned with a unit vector that describes the normal to the plane passing through three neighbors of the considered atom. If the angle $\alpha$ between these vectors at two bonded three-coordinated atoms is such that $|\cos(\alpha)| < 0.7$, i.e. the atoms do not lie on a smooth surface, both of these atoms are supposed to belong to the amorphous domain.

**RESULTS AND DISCUSSION**

The main result of the performed MD simulations is that in all simulation runs initial amorphous carbon clusters transform into hollow shells with sp$^2$ structure. The complete transformation of amorphous carbon clusters into fullerenes within 1 μs is revealed for 8 simulation runs out of 47. Examples of the structure evolution of an amorphous carbon cluster into a fullerene are presented in Figure 1a-h and Figure S2a-h and S3a-h in Supporting Information. As for the rest of the simulation runs, the same transformation process is observed although one or few chains remain attached to the hollow shell with the sp$^2$ structure after 1 μs of the simulation time.

Examples of the time dependences describing evolution of the numbers of different atomic types, numbers of rings and potential energy for several simulation runs in which fullerenes are formed are shown in Figure 1i,j,k and Figure S2i,j,k and S3i,j,k in Supporting Information. Based on these dependences consecutive modifications of the initially amorphous structure can be followed. The initial cluster cut from the bulk amorphous carbon (as described in Supporting Information) contains many dangling bonds and chains attached to the core of the structure (see Figure 1a). These dangling bonds and chains close into polycyclic rings within several ns (see Figure 1b). Then polycyclic rings rearrange into sp$^2$ fragments which consist mainly of 5-, 6-, and 7- rings. Finally, as a result of this first transformation stage a hollow shell with the sp$^2$ structure and several chains attached (usually at both ends) is formed within several tens of nanoseconds (see Figure 1c). The number $N_{3a}$ of three-coordinated atoms which belong to the amorphous domain, the total number $N_r$ of 3-, 4-, 8- and 9-rings and the potential energy $E$ of the system per atom decrease drastically during this stage as shown in Figure 1i,j,k and Figure S2i,j,k and S3i,j,k in Supporting Information. Clusters with the sp$^2$ structure and chains attached typically at both ends[18-21,26] or at one end[22-24] were revealed also in classical MD[18-21] and TBMD[22-24,26] simulations of fullerene formation starting from carbon vapor. Therefore, the results obtained here and discussed below including the information on the last events of chain atom insertion into the forming sp$^2$ shell and defects remaining after the shell formation may be also applicable to fullerene formation in arc discharge and by laser ablation of carbon material. It should be noted that a similar stage of fast formation of the sp$^2$ structure (with the portion of three-coordinated atoms in the system of about 80 – 90 %) was observed also in MD simulations for large carbon systems of several thousands of atoms in the processes of transformation of amorphous carbon[32,33] and nanodiamonds[34] to multishell carbon nanoparticles (onions) and graphitization of nanodiamonds of several nanometers in size[35,36].

At the next transformation stage carbon atoms at the ends of the chains insert into the $sp^2$ structure of the shell. As a result of these atom insertions followed by structure rearrangements the number and length of the chains decrease. The decrease in the number $N_c$ of two-coordinated and one-coordinated carbon atoms in chains during this transformation stage is shown in Figure 1i, Figure 2, and Figure S2i and S3i in Supporting Information. The forming shell tends to be more and more smooth, which corresponds to the increase in the number $N_3$ of three-coordinated atoms belonging to the $sp^2$ network during this transformation stage (see Figure 1i, and Figure S2i and S3i in Supporting Information). The simultaneous decrease in the number $N_c$ of one-coordinated and two-coordinated atoms in chains and the number $N_{3a}$ of three-coordinated atoms which belong to the amorphous domain averaged over the simulation runs (see Figure 2) implies an increase in the number $N_3$ of three-coordinated atoms which belong to the $sp^2$ network averaged over the simulation runs. This confirms formation and smoothing of the $sp^2$ shell in the majority of the simulation runs.

We do not consider here in detail different reactions of atom insertion from long chains into the $sp^2$ structure of the shell. However, we should note that two atoms at both ends of the long chain usually insert at the same time. An example of such a simultaneous two-atom insertion is shown in Figure 4a. Evidently, the suggested division of the transformation process into two stages, formation of the hollow shell of $sp^2$ structure with chains attached to it and chain insertion into the shell is rather speculative because it is difficult to unambiguously determine the moment of cavity formation inside the initial amorphous cluster. Thus we can only roughly estimate that the binding energy increase during the stage of chain insertion ranges from 4 to 11 eV.

In all 8 simulation runs in which all chains got inserted into the shell with $sp^2$ structure, the obtained shell structures contain only 5-, 6- and 7-rings. Thus, according to the definition given above, fullerenes are formed in these 8 simulation runs. One of these fullerenes consists only of three-coordinated atoms and contains only 5- and 6-rings, i.e. has no defects at all (see Figure 1h). As far as we know, this is the first example where a fullerene formed in MD simulations from an initially chaotic system has no defects just after formation. Previously fullerenes without defects have been obtained in MD simulations as a result of subsequent SW reactions[43] or carbon catalyzed rearrangement processes initiated by $C_2$ and $C_3$ attacks[26]. The rest of the fullerenes formed here have structural defects discussed in detail below. The number $N_{3a}$ of three-coordinated atoms which belong to the amorphous domain ranges for the fullerenes formed from 0 (for the fullerene without defects, see Figure 1i) to 4, see Figure 2, and Figure S2i, Figure S3i in Supporting Information, which is clearly much smaller than 30 – 40 atoms of this type in the initial amorphous structure. Minor fluctuations of the number $N_{3a}$ after the fullerene formation are caused by thermal deformation of the fullerene shell without any bond rearrangement.

Therefore, the number $N_{3a}$ can be considered as a measure of the transformation of the amorphous structure into a fullerene.

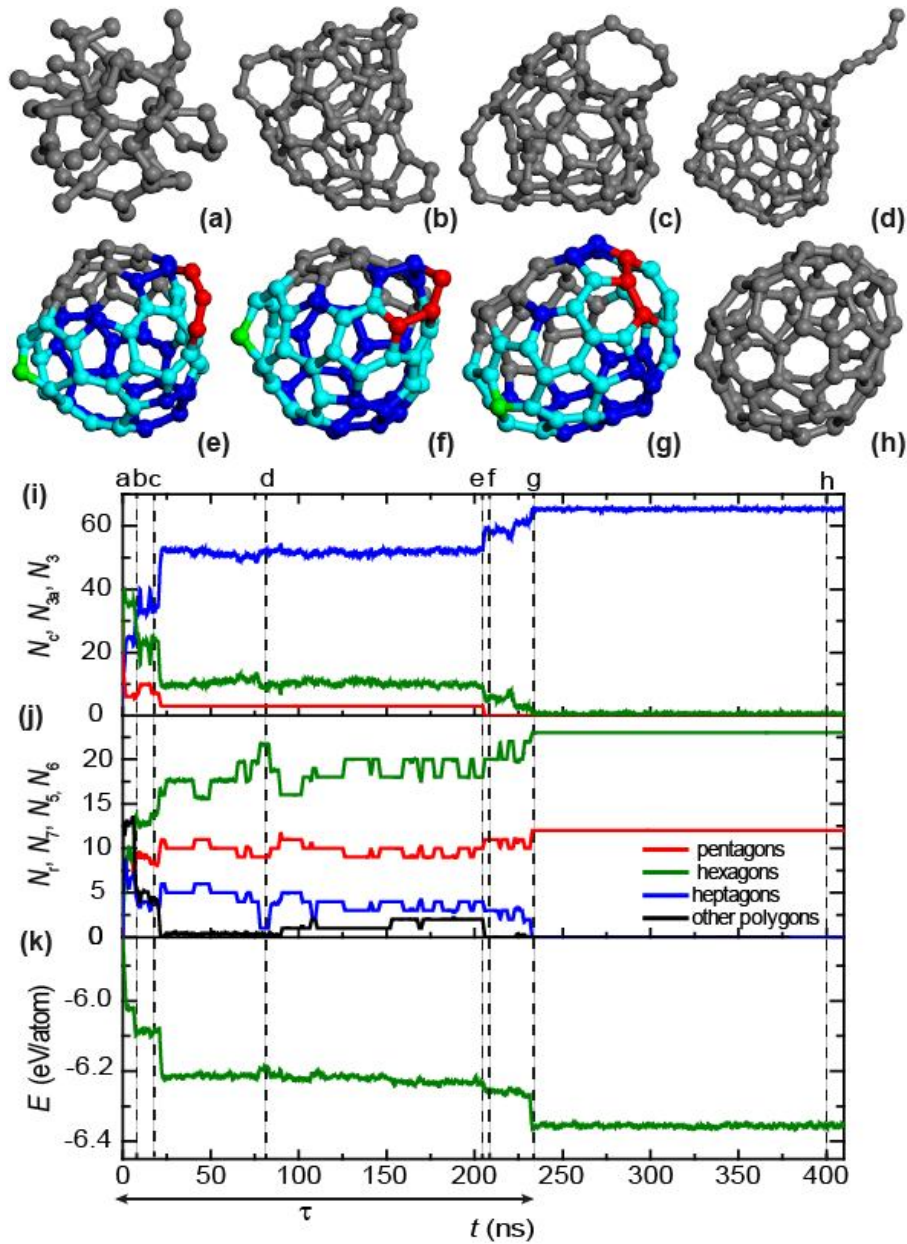

Figure 1. (a-h) Simulated structure evolution in the transformation of the amorphous carbon cluster $C_{66}$ into the fullerene at temperature 2500 K observed at: (a) 0 ns (initial structure), (b) 3 ns, (c) 20 ns, (d) 82.75 ns, (e) 205 ns, (f) 205.3 ns, (g) 232.3 ns and (h) 400 ns. Colored structures (e), (f) and (g) correspond to the last event of chain atom insertion into $sp^2$ network shown schematically in Figure 4f. Carbon atoms belonging to chains, rings of $sp^2$ atoms which contain ≥ 7 atoms and the single two-coordinated atom inserted during this event are colored in red, light blue and light green, respectively. Other atoms shown in scheme (f) of Figure 4 are colored in dark blue. (i) Calculated number of two-

coordinated and one-coordinated carbon atoms in chains, $N_c$ (red line, upper panel), number of three-coordinated atoms belonging to the sp$^2$ network, $N_3$ (blue line, upper panel), number of three-coordinated carbon atoms in the amorphous domain, $N_{3a}$ (green line, upper panel), total number of 5-rings $N_5$ (red line, middle panel), 6-rings $N_6$ (green line, middle panel), 7-rings $N_7$ (blue line, middle panel), other rings $N_r$ (black line, middle panel), potential energy per atom $E$ (green line, lower panel) as functions of time $t$. The calculated values are averaged for each time interval of 1 ns. The moments of time corresponding to structures (a−h) are shown using vertical dashed lines. The transformation time $\tau$ is indicated by the double-headed arrow.

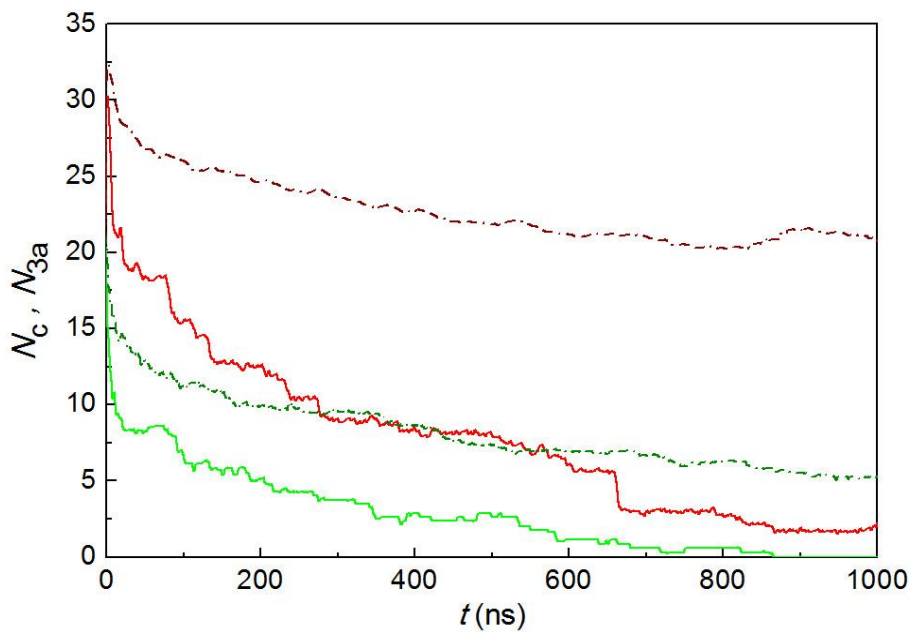

Figure 2. Calculated number of three-coordinated carbon atoms which belong to the amorphous domain, $N_{3a}$, averaged over the simulation runs in which fullerenes are formed and over the rest of the simulation runs (light solid and dark dashed red lines, respectively) as functions of time $t$. Calculated number of two-coordinated and one-coordinated carbon atoms in chains, $N_c$, averaged over simulation runs in which fullerenes are formed and over the rest of the simulation runs (light solid and dark dashed green lines, respectively) as functions of time $t$. The calculated values are averaged for each time interval of 5 ns.

As for the rest of simulation runs in which fullerenes are not formed, the number $N_{3a}$ of three-coordinated atoms which belong to the amorphous domain and the number $N_c$ of atoms in chains keep decreasing during the considered simulation time of 1 μs, see Figure 2. Therefore, the transformation into a fullerene still continues for these simulation runs, and fullerene formation can be expected for the majority of simulations if one waits for a sufficiently long time. However, we cannot exclude the transformation of the initial amorphous carbon cluster into a cluster with another structure corresponding to a local minimum of the free energy for the considered temperature. For example, the study of transformation of carbon clusters with the polycyclic ring structure in the drift tube revealed transformation of these clusters not only into fullerenes but also (for minority of cases) into monocyclic rings[8]. Transformation times of the fullerenes formed in our simulations are listed in Table 2. Since the transformation into the fullerene continues at simulation times of about 1 μs (Figure 2) for majority of the simulation runs the average transformation time at temperature 2500 K exceeds 1 μs.

The majority, 7 out of 8, of the fullerenes formed contain structural defects such as single one-coordinated atoms and 7-rings. The presence of multiple defects in the structure of fullerenes formed is in agreement with other MD simulations of fullerene formation from carbon vapor[17-23,26], graphene flakes[29,30], a short carbon nanotube with open ends[27,28] and a small nanodiamond cluster[31]. Figure 3 shows the defects of the fullerene structure at the moment of fullerene shell formation. Only one fullerene has no defects whereas other 7 fullerenes have from 1 to 5 7-rings. Additionally to 7-rings, 2 fullerenes have a single one-coordinated atom located outside the shell and attached to atoms belonging to two 6-rings. The list of all remaining defects is given in Table 2.

To study possible changes of local structure after the shell formation all 8 simulation runs with the fullerenes formed have been extended for the additional time of $\tau_{add} \sim 0.5 - 1$ μs (this additional time is given in Table 2). Since the fullerenes formed contain several defects, at least several events of bond rearrangement are necessary for healing of all the defects in the fullerene structure. However, neither defect healing nor any other bond rearrangements occur in these fullerenes within the net time of all simulations after the shell formation on the order of 10 μs. Thus defect healing in an isolated fullerene (without the influence of the buffer gas or reactions with carbon vapor) is a slow process at temperature 2500 K with the characteristic time exceeding 10 μs. This result of our simulations is consistent with the quantum chemical calculations that give the high value of the SW reaction activation barrier, 4.8-6.9 eV,[41,45,46,50,70] even in the case where 7-rings are present in the fullerene structure.[50] The opposite result was obtained in previous MD simulations,[43] where a lot of SW reactions were observed at the same temperature of 2500 K for the fullerene with defects and the activation barrier of these SW reactions was estimated to be about 2.5 eV. This difference in the simulation results is related to the use of different versions of the Brenner potential. Whereas the simplified version of the Brenner potential was used in Ref. 43, we use the full version of the Brenner potential with the new set of parameters fitted recently to reproduce the energies of fullerenes and graphene edges[52]. Contrary to the high activation barrier of SW reactions the barrier of $C_2$ insertion is only 1.0-2.5 eV.[47] Thus studies of the reactions of $C_2$ insertion and emission resulting in healing of 7-rings and single two-coordinated atoms and formation of the $sp^2$ structure consisting only of 5-and 6- rings are necessary to understand the last stage of the fullerene formation and the high yield of abundant isomers like the $C_{60}$ fullerene with icosahedral symmetry.

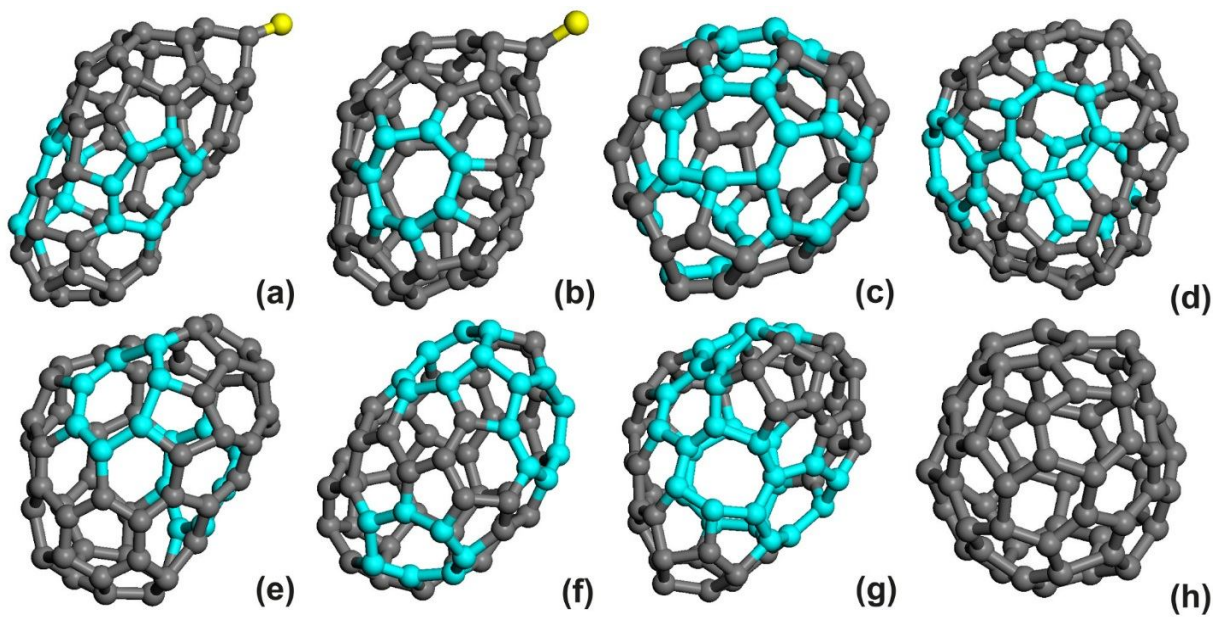

Figure 3. Calculated structure and defects remaining for all fullerenes formed in the molecular dynamic simulations of transformation of amorphous carbon clusters. Single one-coordinated atoms and atoms which belong to 7-rings are coloured in yellow and light blue, respectively. The list of the defects for each fullerene is given in Table 2.

To study the influence of the initial amorphous cluster structure on the fullerene formation, 25 additional simulation runs of 400 ns duration have been performed at temperature 2500 K for the same initial structure. In these simulations the initial structure is the amorphous cluster $C_{66}$ (see Figure 1a) and the difference between the calculations is only in the initial random Maxwell velocity distribution. The $C_{66}$ cluster is the one which transformed into the fullerene without defects during 230 ns in one of 47 main simulation runs discussed above. The time dependences of the numbers $N_{3a}$ and $N_c$ for 47 main simulation runs starting from different initial structures and 25 additional simulation runs starting from the same initial structure are compared in Figure 4S in Supporting Information. Figure 4S shows that the main and additional simulation runs are characterized by similar decrease rates of the number $N_{3a}$ of three-coordinated atoms which belong to the amorphous domain and the number $N_c$ of atoms in chains. However, in none of the additional simulation runs a fullerene is formed. Therefore the transformation of the amorphous cluster into a fullerene is a totally stochastic process, which does not depend on the initial structure of the amorphous cluster. It should be noted that although the duration of the additional simulation runs is about twice greater than the transformation time ~230 ns for the fullerene without defects, this time is more than twice less than the duration of 47 main simulation runs of about 1 μs. Thus a direct comparison of the fullerene yield for the sets of simulation runs with the same and different initial structures is not possible.

Not only reactions of defect healing after the fullerene shell completion but also reactions at the last stage of the formation of the $sp^2$ structure of the fullerene shell are important to understand the atomistic mechanisms of fullerene formation and the origin of the defects remaining after the shell formation. Here the last events of chain insertion into the hollow shell with the $sp^2$ structure are studied in detail. We consider as the last insertion events not only reactions that take place immediately at insertion of chain atoms but also subsequent rearrangement of structure around the place of insertion if this rearrangement is encouraged by the insertion and occurs within few tens of nanoseconds after. Table 2 presents the description of the last insertion events, the binding energy increase, $\Delta E$, and the number of bonds formed and broken during the last insertion event for all the fullerenes formed. 7 different schemes of chain insertion are revealed for 8 fullerenes formed. Such a variety of the insertion schemes is in agreement with the paradigm that the fullerene formation occurs as a stochastic process of the structure self-organization and contradicts fullerene formation mechanisms based on the existence of fullerene precursors of a certain structure at intermediate stages of the fullerene formation (see reviews[1,2] devoted to fullerene formation mechanisms). The schemes of the observed last insertion events are given in Figure 4 and Figure 5S in Supporting Information.

**Table 2.** Characteristics of the last insertion event for all fullerenes formed: correspondence between the figure of the fullerene structure, number of atoms in the amorphous cluster $N_a$ and the scheme of the last insertion event, description of the local structure before and after last insertion of chain atoms, defects remaining after the last insertion event in the area of fullerene structure involved into the event, and number of broken and formed bonds $N_b$ and $N_f$ and binding energy increase $\Delta E$, respectively, during the last insertion event. Calculated parameters for the fullerene structures obtained: list of the defects remaining at the end of each simulation run, transformation time $\tau$ and additional simulation time after the end of the transformation $\tau_{add}$.

| Figure of structure | $N_a$ | Figure of scheme | Structure before insertion | Structure after insertion | Remaining defect | $N_b/N_f$ | $\Delta E$ (eV) | Defects | $\tau$ (ns) | $\tau_{add}$ (ns) |
|---|---|---|---|---|---|---|---|---|---|---|
| Figure 3a | 64 | Figure 4c | 2-atomic chain over 8-ring | 2x6-rings | one-coord. atom | 1/1 | 5.1 | 2 x 7-rings; one-coord.atom | 646 | 1042 |
| Figure 3b | 74 | Figure 4c | 2-atomic chain over 8-ring | 2x6-rings | one-coord. atom | 1/1 | 7.4 | 1 x 7-ring; one-coord. atom | 800 | 520 |
| Figure 3c | 60 | Figure S5c | 2-atomic chain over 9-ring | 3x5-rings & 6-ring | none | 3/3 | 7.2 | 4 x 7-rings | 670 | 766 |
| Figure 3d | 64 | Figure 4b | 2-atomic chain over 10-ring | 5-ring & 2x6-rings | none | 3/4 | 3.8 | 3 x 7-rings | 409 | 749 |
| Figure 3e | 64 | Figure 4e & Figure S5a | 2-atomic chain over 9-ring; | 2x5-rings & 6-ring | none | 12/13 | 6.4 | 2 x 7-rings | 607 | 585 |
| Figure 3f | 62 | Figure 4d | 2-atomic chain over 8- and 7-rings | 2x5-rings, 6-ring & 7-ring | 7-ring | 2/3 | 6.8 | 3 x 7-rings | 134 | 830 |
| Figure 3g | 68 | Figure S5b | 2-atomic chain attached by one end to | 3x5-rings | none | 3/5 | 6.8 | 5 x 7-rings | 123 | 570 |

| | | | 6-atomic ring ; two-coordinated atom; | | | | | | | |
| Figure 3h | 66 | Figure 4f | 3-atomic chain over 6- and 7-rings, two-coordinated atom; | 2x5-rings & 2x6-rings | none | 16/19 | 6.6 | none | 225 | 835 |

The number of fullerenes formed is insufficient to draw quantitative conclusions related to last insertion events at the fullerene shell formation. However, we can find some common features in characteristics of these events. All observed last insertion events are exothermic with the energy release ranging from 3.8 to 7.4 eV (see Table 2). This energy release corresponds to a considerable part (~30-80 %) of the total increase of the binding energy during the chain insertion stage of the transformation.

All last insertion events are represented by the insertion of two- or three-atomic chains attached outside the fullerene shell into the $sp^2$ structure of the shell (see Figure 4 and Figure 5S in Supporting Information). These events occur with rearrangement of the $sp^2$ structure of the shell. For most of the last insertion events (7 out of 8) the inserted chains are attached to atoms of the shell which are members of rings with more than 6 atoms. In these events the rearrangement of the $sp^2$ structure of the shell occurs with transformation of 8-, 9-, and 10-rings into 5-, 6-, and 7-rings, and 4-, 7- and 8-rings into 5-, and 6-rings (see Figure 4 and Figure S5 in Supporting Information).

5 last insertion events take place with only local rearrangement of the shell structure near the atoms to which the chain is attached before the insertion. These events include breaking and formation of 1 – 4 bonds and last up to 1 ns. 3 other last insertion events are accompanied by defect healing (namely healing of the single two-coordinated atom) located rather far from the area of the last insertion event. These events occur with a considerable binding energy increase (more than 6 eV), breaking and formation of up to 20 bonds and last up to several tens of nanoseconds.

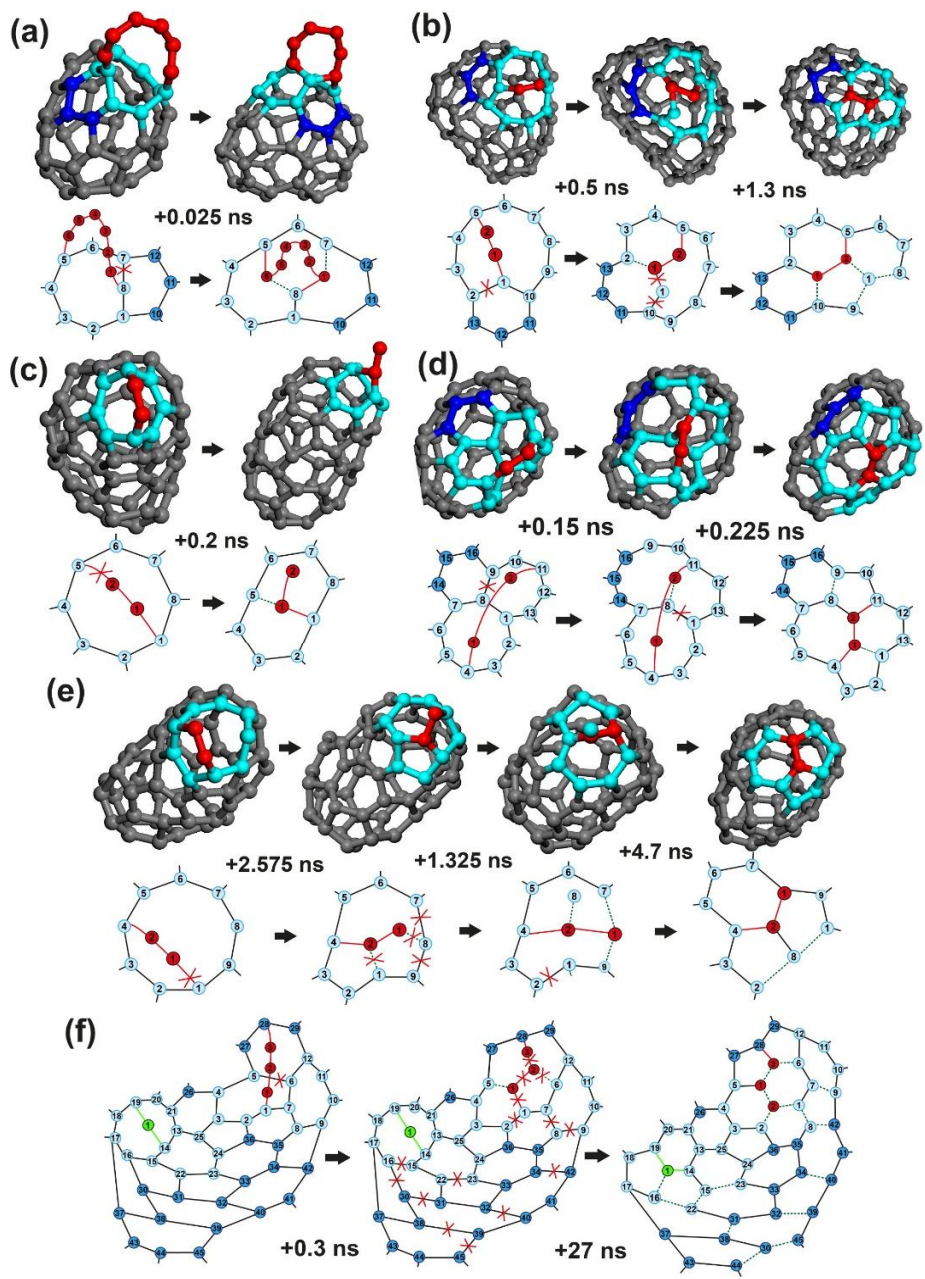

Figure 4. Calculated structures and schemes of insertion of atoms into the sp$^2$ network. (a) Insertion of two atoms from a 6-atom chain (red atoms) at the intermediate stage of fullerene shell formation. (b)-(g) Events of last insertion of atoms, characteristics and description of these events are listed in Table 2. Scheme (f) corresponds to the structures (e), (f) and (g) shown in Figure 1. Scheme (e) shows only a part of the shell area involved into the last insertion event; the complete scheme of this last insertion event is shown in Figure S5a in Supporting Information. To clarify the correspondence between atoms of the structures and in the schemes, atoms corresponding to chains, rings of sp$^2$ atoms which contain ≥ 7 atoms and other atoms in the schemes before the insertion events are coloured in red, light blue and dark blue,

respectively. The time periods between subsequent structures of the same event are indicated. Forming and breaking bonds are indicated by dashed green lines and red crosses, respectively.

For 5 out of 8 last insertion events (including 3 events with healing of single two-coordinated atoms) the $sp^2$ structure which is involved into the event has no local defects after the event. A single one-coordinated atom attached to the atom which is a member of two 7-rings is observed at the last insertion event only as an intermediate state and this one-coordinated atom is incorporated into the $sp^2$ structure with formation of 5- and 6-rings after 5 ns (see Figure 4d and Figure 5Sa in Supporting Information). In contrast to this case, single one-coordinated atoms attached to atoms which are members of two 6-rings (mentioned above among the defects remaining) are considerably more stable and exist at the same temperature 2500 K for times that are at least 3 orders of magnitude longer. We believe that reactions related to the last insertion of chains are of interest for further studies, for example, by *ab initio* methods.

**CONCLUSIONS**

In summary, atomistic modeling of the transformation of amorphous carbon clusters to fullerenes at temperature 2500 K has been performed based on the reactive empirical potential and molecular dynamics method. Two stages of the transformation with different time scales are revealed: 1) formation of a hollow shell with $sp^2$ structure and attached chains with the characteristic time < 100 ns, 2) formation of a fullerene as a result of chain insertion into this hollow shell with $sp^2$ structure with the characteristic time ~ 1 μs. The schemes of reactions of bond rearrangement during the last events of chain insertion into the shell structure, which occur just before the fullerene shell completion, are studied in detail. A considerable variety of these reactions is observed, which means that the performed simulations confirm the paradigm of fullerene formation as a result of self-organization[1,2]. It should be noted that this conclusion is made only for pure carbon systems (or for the systems with the inert buffer gas) and we do not consider here chemical synthesis from polycyclic aromatic hydrocarbon precursor.[71] One of the fullerenes formed consists entirely of three-coordinated atoms arranged in 5- and 6-rings. This example shows that once in a while defect-free carbon cages can be formed from purely amorphous precursors directly without defect healing. The rest of the fullerenes obtained, however, contain local defects of the structure such as 7-rings and single one-coordinated atoms. To observe possible defect healing structural changes have been tracked after the fullerene shell completion at the same time scale as the shell formation. Neither defect healing nor any other events of bond rearrangement have been observed within the net time of all simulations after formation of the fullerene shell on the order of 10 μs. Therefore, processes which lead to selection of abundant isomers (such as $C_{60}$ with icosahedral symmetry) are slow in comparison with the processes of fullerene shell formation.

AUTHOR INFORMATION

**Corresponding Author**

E-mail: liv_ira@hotmail.com

ACKNOWLEDGMENT

This research was supported by the Russian Foundation of Basic Research (Grant 14-02-00739-a). IVL acknowledges Grupos Consolidados del Gobierno Vasco (IT-578-13) and EU-H2020 project "MOSTOPHOS" (n. 646259). We also acknowledge the computational time on the Multipurpose Computing Complex NRC "Kurchatov Institute".

REFERENCES

1. Lozovik, Y. E.; Popov, A. M. Formation and Growth of Carbon Nanostructures: Fullerenes, Nanoparticles, Nanotubes and Cones. *Phys.-Usp.* **1997**, *40*, 717.

2. Irle, S.; Zheng, G. S.; Wang, Z.; Morokuma, K. The $C_{60}$ Formation Puzzle "Solved": QM/MD Simulations Reveal the Shrinking Hot Giant Road of the Dynamic Fullerene Self-Assembly Mechanism. *J. Phys. Chem. B* **2006**, *110*, 14531–14545.

3. Rubin, Y.; Kahr, M.; Knobler, C. B.; Diederich, F.; Wilkins, C. L. The Higher Oxides of Carbon $C_{8n}O_{2n}$ (N = 3-5): Synthesis, Characterization, and X-Ray Crystal Structure. Formation of Cyclo[N]Carbon Ions $C_{n+}$ (N = 18, 24), $C_{n-}$ (N = 18, 24, 30), and Higher Carbon Ions Including $C^{+}_{60}$ in Laser Desorption Fourier Transform Mass Spectrometric Experiments. *J. Am. Chem. Soc.* **1991**, *113*, 495–500.

4. McElvany, S. W.; Ross, M. M.; Goroff, N. S.; Diederich, F. Cyclocarbon Coalescence: Mechanisms for Tailor-Made Fullerene Formation. *Science* **1993**, *259*, 1594–1596.

5. von Helden, G.; Gotts, N. G.; Bowers, M.T. Experimental Evidence for the Formation of Fullerenes by Collisional Heating of Carbon Rings in the Gas Phase. *Nature* **1993**, *363*, 60–63.

6. Hunter, J.; Fye, J.; Jarrold, M. F. Carbon Rings. *J. Phys. Chem.* **1993**, *97*, 3460–3462.

7. Hunter, J.; Fye, J.; Jarrold, M. F. Annealing and Dissociation of Carbon Rings. *J. Chem. Phys.* **1993**, *99*, 1785–1795.

8. Hunter, J.; Fye, J.; Jarrold, M. F. Annealing $C_{60}^{+}$: Synthesis of Fullerenes and Large Carbon Rings. *Science* **1993**, *260*, 784–786.

9. Yeretzian, C.; Hansen, K.; Diederich, A. F.; Whetten, R. L. Coalescence Reactions of Fullerenes. *Nature* **1992**, *359*, 44–47.

10. Chuvilin, A.; Kaiser, U.; Bichoutskaia, E.; Besley, N. A.; Khlobystov, A. N. Direct Transformation of Graphene to Fullerene. *Nat. Chem.* **2010**, *2*, 450–453.

11. Sinitsa, A. S.; Chamberlain, T. W.; Zoberbier, T; Lebedeva, I. V.; Popov, A. M.; Knizhnik, A. A.; McSweeney, R. L.; Biskupek, J.; Kaiser, U.; Khlobystov, A. N. Formation of Nickel Clusters Wrapped in Carbon Cages: Toward New Endohedral Metallofullerene Synthesis. *Nano Lett.* **2017**, *17*, 1082–1089.

12. Jones, R. O. Density Functional Study of Carbon Clusters $C_{2n}$ (2≤N≤16). I. Structure and Bonding in the Neutral Clusters. *J. Chem. Phys.* **1999**, *110*, 5189–5200.

13. Killblane, C.; Gao, Y.; Shao N.; Zeng, X. C. Search for Lowest-Energy Nonclassical Fullerenes III: $C_{22}$. *J. Phys. Chem. A* **2009**, *113*, 8839–8844.

14. Portmann, S.; Galbraith, J. M.; Schaefer, H. F.; Scuseria, G. E.; Lüthi, H. P. Some New Structure of $C_{28}$. *Chem. Phys. Lett.* **1999**, *301*, 98–104.

15. Barnard, A. S.; Russo, S. P.; Snook, I. K. Size Dependent Phase Stability of Carbon Nanoparticles: Nanodiamond versus Fullerenes. *J. Chem. Phys.* **2003**, *118*, 5094–5097.

16. Jiang, Q.; Chen, Z. P. Thermodynamic Phase Stabilities of Nanocarbon. *Carbon* **2006**, *44*, 79–83.

17. Chelikowsky, J. R. Formation of $C_{60}$ Clusters via Langevin Molecular Dynamics. *Phys. Rev. B* **1992**, *45*, 12062–12070.

18. Makino, S.; Oda, T.; Hiwatari, Y. Classical Molecular Dynamics for the Formation Process of a Fullerene Molecule. *J. Phys. Chem. Solids* **1997**, *58*, 1845–1851.

19. László, I. Formation of Cage-Like $C_{60}$ Clusters in Molecular-Dynamics Simulations. *Europhys. Lett.* **1998**, *44*, 741–746.

20. Yamaguchi, Y.; Maruyama, S. A Molecular Dynamics Simulation of the Fullerene Formation Process. *Chem. Phys. Lett.* **1998**, *286*, 336–342.

21. Yamaguchi, Y.; Maruyama, S. A Molecular Dynamics Study on the Formation of Metallofullerene. *Eur. Phys. J. D* **1999**, *9*, 385–388.

22. Irle, S.; Zheng, G. S.; Elstner, M.; Morokuma, K. From $C_2$ Molecules to Self-Assembled Fullerenes in Quantum Chemical Molecular Dynamics. *Nano Lett.* **2003**, *3*, 1657–1664.

23. Zheng, G. S.; Irle, S.; Morokuma, K. Towards Formation of Buckminsterfullerene $C_{60}$ in Quantum Chemical Molecular Dynamics. *J. Chem. Phys.* **2005**, *122*, 014708.

24. Yamaguchi, Y.; Colombo, L.; Piseri, P.; Ravagnan, L.; Milani, P. Growth of sp−$sp^2$ Nanostructures in a Carbon Plasma. *Phys. Rev. B* **2007**, *76*, 134119.

25. Hussien, A.; Yakubovich, A. V.; Solov'yov, A. V.; Greiner, W. Phase Transition, Formation and Fragmentation of Fullerenes. *Eur. Phys. J. D* **2010**, *57*, 207–217.

26. Qian, H.-J.; Wang, Y.; Morokuma, K. Quantum Mechanical Simulation Reveals the Role of Cold Helium Atoms and the Coexistence of Bottom-Up and Top-Down Formation Mechanisms of Buckminsterfullerene from Carbon Vapor. *Carbon* **2017**, *114*, 635-641.

27. Irle, S.; Zheng, G. S.; Elstner, M.; Morokuma, K., Formation of Fullerene Molecules from Carbon Nanotubes: A Quantum Chemical Molecular Dynamics Study. *Nano Lett.* **2003**, *3*, 465–470.

28. Zheng, G. S.; Irle, S.; Elstner, M.; Morokuma, K. Quantum Chemical Molecular Dynamics Model Study of Fullerene Formation from Open-Ended Carbon Nanotubes. *J. Phys. Chem. A* **2004**, *108*, 3182–3194.

29. Lebedeva, I. V.; Knizhnik, A. A.; Bagatur'yants, A. A.; Potapkin, B. V. Kinetics of 2D–3D Transformations of Carbon Nanostructures. *Physica E* **2008**, *40*, 2589–2595.

30. Lebedeva, I. V.; Knizhnik, A. A.; Popov, A. M.; Potapkin, B. V. Ni-Assisted Transformation of Graphene Flakes to Fullerenes. *J. Phys. Chem. C* **2012**, *116*, 6572–6584.

31. Lee, G.-D.; Wang, C. Z.; Yu, J.; Yoon, E.; Ho, K. M. Heat-Induced Transformation of Nanodiamond into a Tube-Shaped Fullerene: A Molecular Dynamics Simulation. *Phys. Rev. Lett.* **2003**, *91*, 265701.

32. Lau, D. W. M.; McCulloch, D. G.; Marks, N. A.; Madsen, N. R.; Rode, A. V. High-Temperature Formation of Concentric Fullerene-Like Structures within Foam-Like Carbon: Experiment and Molecular Dynamics Simulation. *Phys. Rev. B* **2007**, *75*, 233408.

33. Powles, R. C.; Marks, N. A.; Lau, D. W. M. Self-Assembly of sp$^2$-bonded Carbon Nanostructures from Amorphous Precursors. *Phys. Rev. B* **2009**, *79*, 075430.

34. Los, J. H.; Pineau, N.; Chevrot, G.; Vignoles, G.; Leyssale, J.-M. Formation of Multiwall Fullerenes from Nanodiamonds Studied by Atomistic Simulations. *Phys. Rev. B* **2009**, *80*, 155240.

35. Leyssale, J.-M.; Vignoles, G. L. Molecular Dynamics Evidences of the Full Graphitization of a Nanodiamond Annealed at 1500 K. *Chem. Phys. Lett.* **2008**, *454*, 299–304.

36. Bródka, A.; Zerda, T.W.; Burian, A. Graphitization of Small Diamond Cluster – Molecular Dynamics Simulation. *Diamond and Related Materials*, **2006**, *15*, 1818–1821.

37. Lozovik, Yu. E.; Popov, A. M. Role of Reactions of Molecule C$_2$ Insertion and Emission in Relative Abundances of Fullerenes and Their Isomers. *Mol. Mater.* **1998**, *10*, 83–86.

38. Krestinin, A. V.; Moravsky, A. P. Mechanism of Fullerene Synthesis in the Arc Reactor. *Chem. Phys. Lett.* **1998**, *286*, 479–484.

39. Yi, J.-Y.; Bernholc, J. Reactivity, Stability, and Formation of Fullerenes. *Phys. Rev. B* **1992**, *48*, 5724–5727.

40. Xu, C.; Scuseria, G. E. Tight-Binding Molecular Dynamics Simulations of Fullerene Annealing and Fragmentation. *Phys. Rev. Lett.* **1994**, *72*, 669–672.

41. Eggen, B. R.; Heggie, M. I.; Jungnickel, G.; Latham, C. D.; Jones, R.; Briddon, P. R. Autocatalysis During Fullerene Growth. *Science* **1996**, *272*, 87–89.

42. Marcos, P. A.; López, M. J.; Rubio, A.; Alonso J. A. Thermal Road for Fullerene Annealing. *Chem. Phys. Lett.* **1997**, *273*, 367–370.

43. Maruyama, S.; Yamaguchi, Y. A Molecular Dynamics Demonstration of Annealing to a Perfect C$_{60}$ Structure. *Chem. Phys. Lett.* **1998**, *286*, 343–349.

44. Osawa, E.; Ueno, H.; Yoshida, M.; Slanina, Z.; Zhao, X.; Nishiyama, M.; Saito, H. Combined Topological and Energy Analysis of the Annealing Process in Fullerene Formation. Stone-Wales Interconversion Pathways among IPR Isomers of Higher Fullerenes. *J. Chem. Soc. Perkin Trans. 2* **1998**, *4*, 943–950.

45. Ewels, C. P.; Heggie, M. I.; Briddon, P. R. Adatoms and Nanoengineering of Carbon. *Chem. Phys. Lett.* **2002**, *351*, 178–182.

46. Bettinger, H. F.; Yakobson, B. I.; Scuseria, G. E. Scratching the Surface of Buckminsterfullerene: The Barriers for Stone-Wales Transformation through Symmetric and Asymmetric Transition States. *J. Am. Chem. Soc.* **2003**, *125*, 5572–5580.

47. Dang, J.-S.; Wang, W.-W.; Zheng, J.-J.; Zhao, X.; Osawa, E.; Nagase, S. Fullerene Genetic Code: Inheritable Stability and Regioselective $C_2$ Assembly. *J. Phys. Chem. C* **2012**, *116*, 16233–16239.

48. Wang, W.-W.; Dang, J.-S.; Zheng, J.-J.; Zhao, X.; Nagase, S. Selective Growth of Fullerenes from $C_{60}$ to $C_{70}$: Inherent Geometrical Connectivity Hidden in Discrete Experimental Evidence. *J. Phys. Chem. C* **2013**, *117*, 2349–2357.

49. Mitchell, D.; Fowler, P. W.; Zerbetto, F. A Generalized Stone–Wales Map: Energetics and Isomerizations of $C_{40}$ Carbon Cages. *J. Phys. B: At. Mol. Opt. Phys.* **1996**, *29*, 4895–4906.

50. Osawa, E.; Honda, K. Stone-Wales Rearrangements Involving Heptagonal Defects. *Fullerene Sci. Technol.* **1996**, *4*, 939–961.

51. Brenner, D. W. Empirical Potential for Hydrocarbons for Use in Simulating the Chemical Vapor Deposition of Diamond Film. *Phys. Rev. B* **1990**, *42*, 9458−9471.

52. Sinitsa, A. S.; Lebedeva, I. V.; Knizhnik, A. A.; Popov, A. M., Skowron, S. T.; Bichoutskaia, E. Formation of Nickel–Carbon Heterofullerenes under Electron Irradiation. *Dalton Trans.* **2014**, *43*, 7499–7513.

53. Brenner, D. W.; Shenderova, O. A.; Harrison, J. A.; Stuart, S. J.; Ni, B.; Sinnott, S. B. A Second-Generation Reactive Empirical Bond Order (REBO) Potential Energy Expression for Hydrocarbons. *J. Phys.: Condens. Matter* **2002**, *14*, 783–802.

54. van Duin, A. C. T.; Dasgupta, S.; Lorant, F.; Goddard, W. A. ReaxFF: A Reactive Force Field for Hydrocarbons. *J. Phys. Chem. A* **2001**, *105*, 9396–9409.

55. Chenoweth, K.; van Duin, A. C. T.; Goddard, W. A. ReaxFF Reactive Force Field for Molecular Dynamics Simulations of Hydrocarbon Oxidation. *J. Phys. Chem. A* **2008**, *112*, 1040–1053.

56. Srinivasan, S. G.; van Duin, A. C. T.; Ganesh, P. Development of a ReaxFF Potential for Carbon Condensed Phases and Its Application to the Thermal Fragmentation of a Large Fullerene. *J. Phys. Chem. A* **2015**, *119*, 571–580.

57. Gan, C. K.; Srolovitz, D. J. First-Principles Study of Graphene Edge Properties and Flake Shapes. *Phys. Rev. B*, **2010**, *81*, 125445.

58. Ivanovskaya, V. V.; Zobelli, A.; Wagner, P.; Heggie, M. I.; Briddon, P. R.; Rayson, M. J.; Ewels, C. P. Low-Energy Termination of Graphene Edges via the Formation of Narrow Nanotubes. *Phys. Rev. Lett.*, **2011**, *107*, 065502.

59. Kroes, J. M. H.; Akhukov, M. A.; Los, J. H.; Pineau, N.; Fasolino, A. Mechanism and Free-Energy Barrier of the Type-57 Reconstruction of the Zigzag Edge of Graphene. *Phys. Rev. B*, **2011**, *83*, 165411.

60. Wassmann, T.; Seitsonen, A. P.; Saitta, A. M.; Lazzeri, M.; Mauri, F. Structure, Stability, Edge States, and Aromaticity of Graphene Ribbons. *Phys. Rev. Lett.*, **2008**, *101*, 096402.

61. Perdew, J. P.; Zunger, A. Self-Interaction Correction to Density-Functional Approximations for Many-Electron Systems. *Phys. Rev. B* **1981**, *23*, 5048–5079.

62. Perdew, J. P.; Burke, K.; Ernzerhof, M. Generalized Gradient Approximation Made Simple. *Phys. Rev. Lett.* **1996**, *77*, 3865–3868.

63. Skowron, S. T.; Lebedeva, I. V.; Popov, A. M.; Bichoutskaia, E. Energetics of Atomic Scale Structure Changes in Graphene. *Chem. Soc. Rev.* **2015**, *44*, 3143–3176.

64. Lebedeva, I. V.; Chamberlain, T. W.; Popov, A. M.; Knizhnik, A. A.; Zoberbier, T.; Biskupek, J.; Kaiser, U.; Khlobystov, A. N. The Atomistic Mechanism of Carbon Nanotube Cutting Catalyzed by Nickel under an Electron Beam. *Nanoscale* **2014**, *6*, 14877–14890.

65. Knizhnik, A. A.; MD-kMC code. *Kintech Lab* **2000-2016**.

66. Swope, W. C.; Andersen, H. C.; Berens, P. H.; Wilson, K. R. A Computer Simulation Method for the Calculation of Equilibrium Constants for the Formation of Physical Clusters of Molecules: Application to Small Water Clusters. *J. Chem. Phys.* **1982**, *76*, 637–649.

67. Allen, M. P.; Tildesley, D. J. *Computer Simulation of Liquids*; Clarendon Press: Oxford, 1987.

68. Berendsen, H. J. C.; Postma, J. P. M.; van Gunsteren, W. F.; DiNola, A.; Haak, J. R. Molecular Dynamics with Coupling to an External Bath. *J. Chem. Phys.* **1984**, *81*, 3684.

69. Franzblau, D. C. Computation of Ring Statistics for Network Models of Solids. *Phys. Rev. B,* **1991**, *44*, 4925.

70. Kabir, M.; Mukherjee, S.; Saha-Dasgupta, T. Substantial Reduction of Stone-Wales Activation Barrier in Fullerene. *Phys. Rev. B* **2011**, *84*, 205404.

71. Scott, L. T.; Boorum, M. M.; McMahon, B. J.; Hagen, S.; Mack J.; Blank J.; Wegner, H.; de Meijere, A. A Rational Chemical Synthesis of $C_{60}$. *Science* **2002**, *95*, 1500–1503.

# Supporting Information for

# "Transformation of Amorphous Carbon Clusters to Fullerenes"

*Alexander S. Sinitsa[1], Irina V. Lebedeva[2*], Andrey M. Popov[3], Andrey A. Knizhnik[4]*

[1] *National Research Centre "Kurchatov Institute", Kurchatov Square 1, Moscow 123182, Russia*
[2] *Nano-Bio Spectroscopy Group and ETSF, Universidad del País Vasco, CFM CSIC-UPV/EHU, San Sebastian 20018, Spain*
[3] *Institute for Spectroscopy of Russian Academy of Sciences, Fizicheskaya Street 5, Troitsk, Moscow 108840, Russia*
[4] *Kintech Lab Ltd., 3rd Khoroshevskaya Street 12, Moscow 123298, Russia*

**Corresponding Author**

E-mail: liv_ira@hotmail.com

**Generation of the initial structure of amorphous carbon clusters**

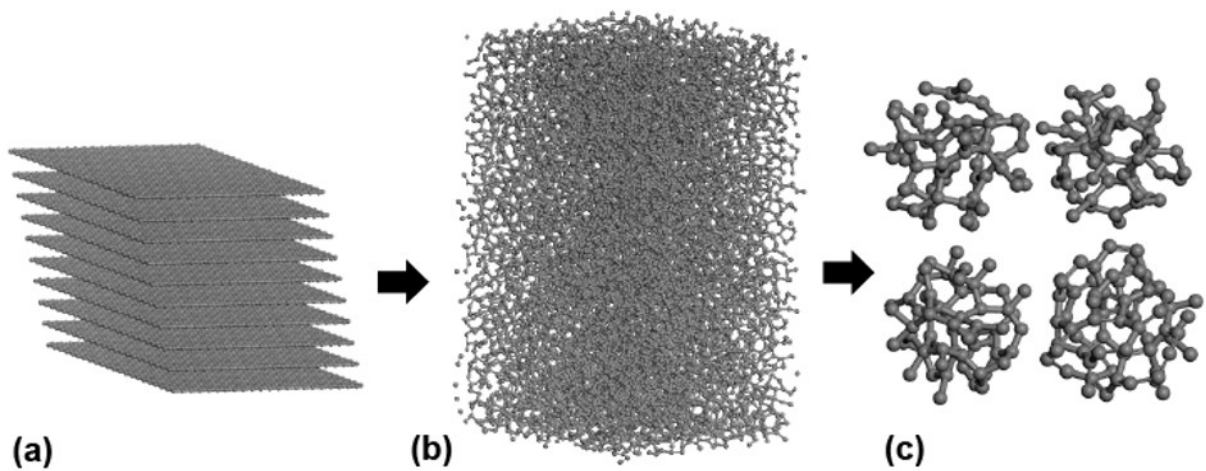

Figure S1. Scheme of generation of the amorphous carbon structure: (a) 10 graphite layers of 800 atoms before annealing; (b) amorphous carbon after annealing at temperature 8000 K and quenching to room temperature; (c) examples of amorphous carbon clusters cut from the amorphous structure obtained.

There are two common methods for preparation of amorphous carbon structures for molecular dynamics simulations: (a) random generation or sequential addition of carbon dimers (see, for example, ref 72) or (b) annealing of several graphite layers or diamond structure at high (6000-7000 K) temperature followed by quenching to room temperature[73-77]. Method (a) is closer to carbon plasma modeling for a variety of reactors, and formation of fullerenes in such a carbon plasma was studied in detail (see review 2). Here we are interested in self-organization of already prepared amorphous carbon clusters, so in our work we use method (b) for setting the initial structure of amorphous carbon.

We performed MD annealing of ten graphite layers (8000 carbon atoms) structure at 8000 K, followed by rapid (10 K/fs) quenching to room temperature. The amorphous carbon cluster was cut from the obtained amorphous structure using the following algorithm: firstly, the "center" of the cluster was chosen randomly in the quenched amorphous carbon bulk, then atoms farther than a predetermined distance from this chosen "center" (4.5 Å for clusters in our simulations) were removed to obtain a ball of amorphous carbon. If the total number of atoms was odd, one atom was deleted, so all clusters contain only even number of atoms.

**Analysis of structure evolution during fullerene formation**

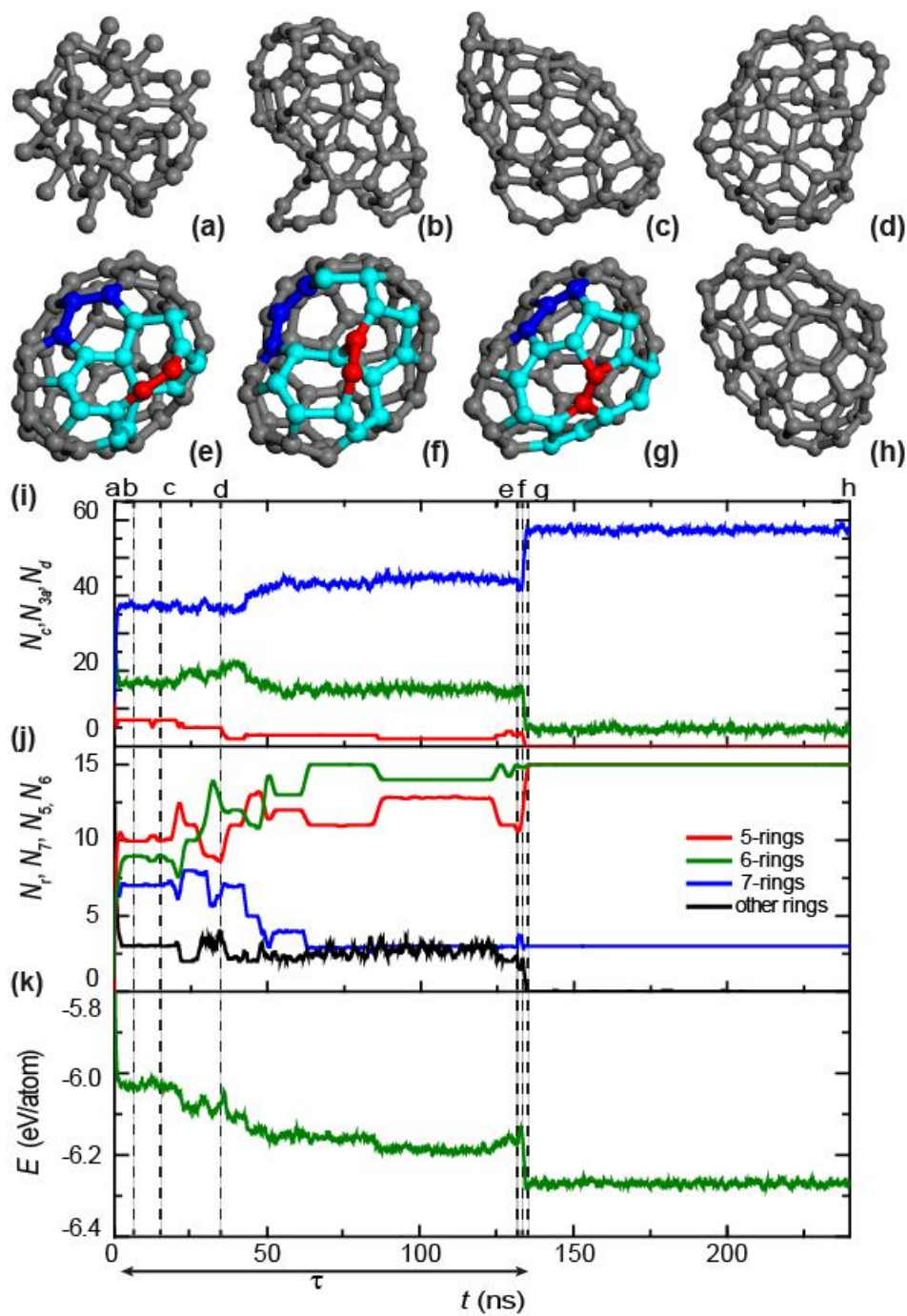

Figure S2. (a-h) Simulated structure evolution in the transformation of the amorphous carbon cluster $C_{62}$ into the fullerene at temperature 2500 K observed at: (a) 0 ns (initial structure), (b) 5 ns, (c) 20 ns, (d) 40 ns, (e) 133.625 ns, (f) 133.775 ns, (g) 134 ns and (h) 625 ns. Colored structures (e), (f) and (g) correspond to the last event of chain atom insertion into the $sp^2$ network shown schematically in Figure 4d. Carbon atoms belonging to chains and rings of $sp^2$ atoms which contain $\geq 7$ atoms are colored in red and light blue respectively. Other atoms shown in scheme (d) of Figure 4 are colored in dark blue. (i) Calculated number of two-coordinated and one-coordinated atoms in chains, $N_c$ (red line, upper panel), number of three-coordinated atoms belonging to the $sp^2$ network, $N_3$ (blue line, upper panel), number of three-coordinated atoms in the amorphous domain, $N_{3a}$ (green line, upper panel), total number of 5-rings $N_5$ (red line, middle panel), 6-rings $N_6$ (green line, middle panel), 7-rings $N_7$ (blue line, middle panel),

other rings $N_r$ (black line, middle panel), potential energy per atom $E$ (green line, lower panel) as functions of time $t$. The calculated values are averaged for each time interval of 1 ns. The moments of time corresponding to structures (a−h) are shown using vertical dashed lines. The transformation time $\tau$ is indicated by the double-headed arrow.

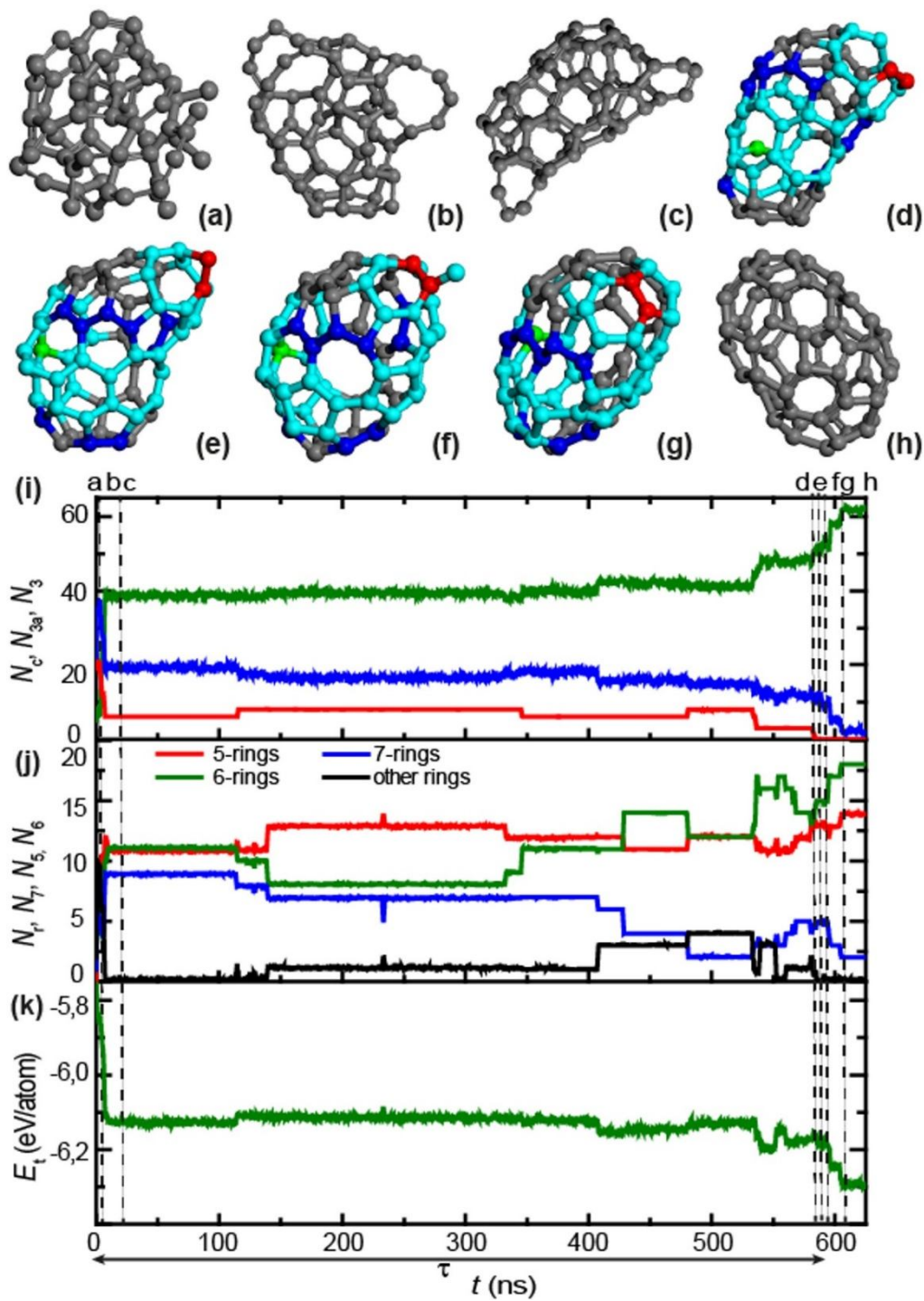

Figure S3. (a-h) Simulated structure evolution in the transformation of the amorphous carbon cluster $C_{64}$ into the fullerene at temperature 2500 K observed at: (a) 0 ns (initial structure), (b) 6 ns, (c) 20 ns, (d) 584.2 ns, (e) 586.575 ns, (f) 595.9 ns, (g) 604.6 ns and (h) 625 ns. Colored structures (d), (e), (f) and (g) correspond to the last event of chain atom insertion into the $sp^2$ network shown schematically in Figure S5a. Carbon atoms belonging to chains, rings of $sp^2$ atoms which contain ≥ 7 atoms and single two-

coordinated atom inserted during this event are colored in red, light blue and light green, respectively. Other atoms shown in scheme (a) of Figure S5 are colored in dark blue. (i) Calculated number of two-coordinated and one-coordinated atoms in chains, $N_c$ (red line, upper panel), number of three-coordinated atoms belonging to the sp$^2$ network, $N_3$ (blue line, upper panel), number of three-coordinated atoms in the amorphous domain, $N_{3a}$ (green line, upper panel), total number of 5-rings $N_5$ (red line, middle panel), 6-rings $N_6$ (green line, middle panel), 7-rings $N_7$ (blue line, middle panel), other rings $N_r$ (black line, middle panel), potential energy per atom $E_t$ (green line, lower panel) as functions of time $t$. The calculated values are averaged for each time interval of 1 ns. The moments of time corresponding to structures (a−h) are shown using vertical dashed lines. The transformation time $\tau$ is indicated by the double-headed arrow.

**Analysis of simulation runs starting from the same initial structure**

To study the influence of the initial structure of amorphous cluster on the fullerene formation, 25 additional simulation runs duration of 400 ns are performed at 2500 K temperature for the same initial structure. This set of simulation runs is performed for the initial structure of the amorphous cluster $C_{66}$ (see Figure 1a). This is the structure that transformed into the fullerene without any defects within 230 ns in one of 47 simulation runs starting from different initial structures. However, in none of these additional simulation runs fullerene formation is revealed, while the similar decrease rate of the number $N_{3a}$ of three-coordinated atoms which belong to the amorphous domain and the number $N_c$ of atoms in chains is observed. The time dependences of the numbers $N_{3a}$ and $N_c$ for 47 simulation runs starting from different initial structures and 25 simulation runs starting from the same initial structure are compared in Figure S4. Therefore, the transformation of amorphous carbon clusters to fullerenes is a totally stochastic process, which does not depend on the initial structure of the amorphous cluster.

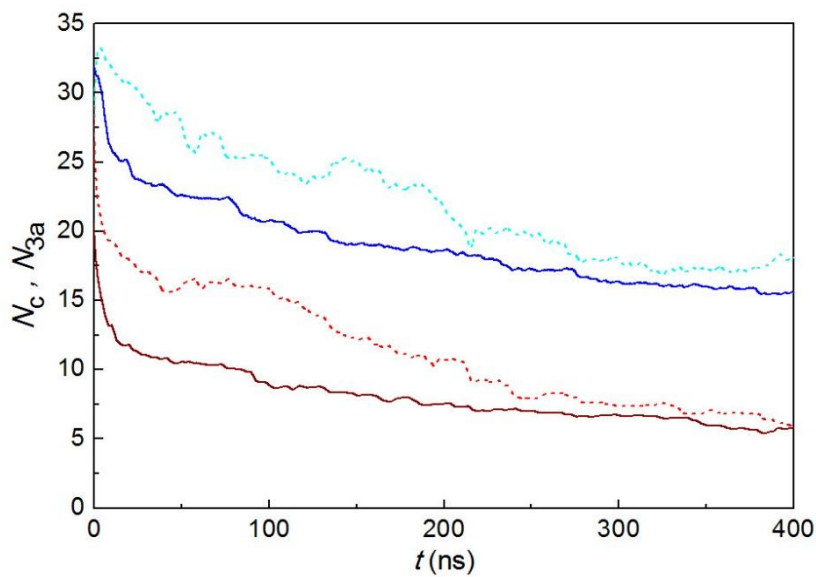

Figure S4. Calculated number of three-coordinated carbon atoms which belong to the amorphous domain, $N_{3a}$, averaged over 47 simulation runs starting from different initial structures and 25 simulation runs starting from the same initial structure (dark solid and light dashed blue lines, respectively) as functions of time $t$. Calculated number of two-coordinated and one-coordinated carbon atoms in chains, $N_c$, averaged over 47 simulation runs starting from different initial structures and 25 simulation runs starting from the same initial structure (dark solid and light dashed red lines, respectively) as functions of time $t$. The calculated values are averaged for each time interval of 5 ns.

**Analysis of last chain insertion events**

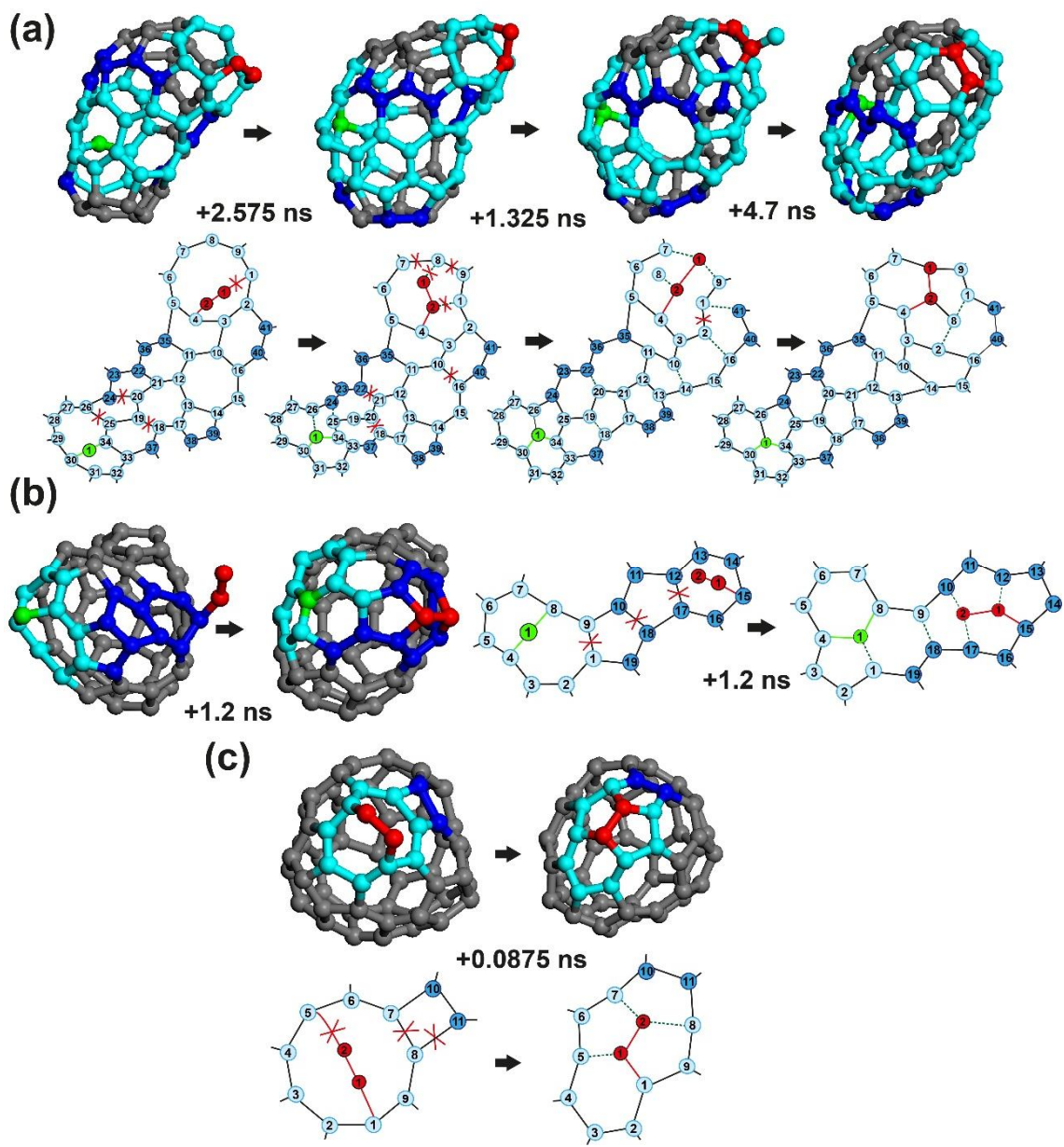

Figure S5. Calculated structures and schemes for events (a-c) of last insertion of atoms in the sp$^2$ network. The characteristics and description of these events are listed in Table 1. Schemes (a) correspond to structures (d-g) in Figure S3. To clarify the correspondence between atoms of the structures and in the schemes, atoms belonging to chains before the event, rings of sp$^2$ atoms which contain ≥ 7 atoms before the event and other atoms on the schemes are coloured in red, light blue and dark blue, respectively. The time periods between subsequent structures corresponding to the same event are indicated. Forming and breaking bonds are indicated by dashed green lines and red crosses, respectively.

**Calculated characteristics for all simulation runs**

The complete transformation of amorphous carbon clusters into fullerenes within 1 μs has been revealed for 8 simulation runs out of 47. In all these 8 simulation runs all chains are finally inserted into the shell with sp$^2$ structure, so no two-coordinated and one-coordinated atoms in chains are observed after the fullerene formation ($N_c^{fin}$ = 0). Minor fluctuations in the number of three-coordinated atoms in the amorphous domain $N_{3a}$ ($N_{3a}$ ~ 0 – 4) after the fullerene formation are caused by thermal deformation of the fullerene shell without any bond rearrangement. As for the rest of the simulation runs in which fullerenes are not formed, the final structure usually looks like a hollow shell with one to three dangling carbon chains (very similar to, for example, structures c,d,e on Figure S2). So both the number of two-coordinated and one-coordinated atoms in chains and the number of three-coordinated atoms in the amorphous domain at the end of each simulation run ($N_c$ and $N_{3a}$) are rather small (usually less than 10 atoms for each) and continue decreasing (see Figure 2). Therefore, the transformation into the fullerene still continues in these simulation runs, and the fullerene formation can be expected for the majority of the simulations if one waits for a sufficiently long time.

Table S1. Calculated parameters for all simulation runs: calculation number, number of atoms in the amorphous cluster $N_a$, calculation time $\tau_{run}$, number of two-coordinated and one-coordinated atoms in chains $N_c^{fin}$ and number of three-coordinated atoms in the amorphous domain $N_{3a}^{fin}$ remaining at the end of each simulation run.

| Calculation number | $N_a$ | $\tau_{run}$ (ns) | $N_c^{fin}$ | $N_{3a}^{fin}$ |
|---|---|---|---|---|
| 1 | 60 | 1038 | 4 | 10 |
| 2 | 64 | 1012 | 6 | 8 |
| 3 | 62 | 1031 | 6 | 4 |
| 4 | 66 | 1045 | 10 | 4 |
| 5 | 66 | 1020 | 4 | 4 |
| 6 | 68 | 1063 | 4 | 6 |
| 7 | Fullerene, see Figure 3g/Table 2 row 7 | | | |
| 8 | 60 | 1091 | 6 | 6 |
| 9 | 66 | 1007 | 6 | 1 |
| 10 | Fullerene, see Figure 3a/Table 2 row 1 | | | |
| 11 | 66 | 1026 | 6 | 4 |
| 12 | 64 | 1026 | 2 | 2 |
| 13 | 70 | 1071 | 8 | 13 |
| 14 | 58 | 1045 | 4 | 1 |
| 15 | 62 | 1132 | 4 | 1 |
| 16 | 66 | 1173 | 8 | 2 |
| 17 | 64 | 1002 | 6 | 6 |
| 18 | Fullerene, see Figure 3c/Table 2 row 3 | | | |
| 19 | 68 | 1187 | 4 | 1 |
| 20 | 66 | 1167 | 10 | 2 |
| 21 | 66 | 1041 | 8 | 4 |
| 22 | 64 | 1023 | 4 | 4 |
| 23 | 64 | 1004 | 8 | 4 |
| 24 | 60 | 1100 | 2 | 2 |
| 25 | 62 | 1060 | 4 | 2 |
| 26 | 60 | 1024 | 2 | 2 |
| 27 | 64 | 1047 | 8 | 3 |
| 28 | 60 | 1018 | 6 | 2 |
| 29 | 64 | 1055 | 6 | 4 |
| 30 | 62 | 1022 | 2 | 4 |
| 31 | Fullerene, see Figure 3h/Table 2 row 8 | | | |
| 32 | 64 | 1112 | 6 | 6 |
| 33 | 74 | 1000 | 4 | 3 |
| 34 | 64 | 1060 | 6 | 4 |
| 35 | Fullerene, see Figure 3f/Table 2 row 6 | | | |
| 36 | 74 | 1124 | 2 | 4 |
| 37 | 62 | 1100 | 4 | 6 |
| 38 | Fullerene, see Figure 3b/Table 2 row 2 | | | |
| 39 | 74 | 1115 | 2 | 6 |
| 40 | 62 | 1172 | 4 | 1 |
| 41 | 58 | 1034 | 8 | 2 |
| 42 | 58 | 1068 | 2 | 1 |
| 43 | 60 | 1085 | 10 | 1 |
| 44 | Fullerene, see Figure 3e/Table 2 row 5 | | | |
| 45 | 60 | 1048 | 2 | 4 |
| 46 | 64 | 1160 | 2 | 1 |
| 47 | Fullerene, see Figure 3d/Table 2 row 4 | | | |

**References**

72. Irle, S.; Zheng, G. S.; Wang, Z.; Morokuma, K. The $C_{60}$ Formation Puzzle "Solved": QM/MD Simulations Reveal the Shrinking Hot Giant Road of the Dynamic Fullerene Self-Assembly Mechanism. *J. Phys. Chem. B* **2012**, *110*, 14531.

73. Sinnott, S. B.; Coltor, R. J.; White, C. T.; Shenderova, O. A.; Brenner, D. W.; Harrison, J. A. Atomistic Simulations of the Nanometer-Scale Indentation of Amorphous-Carbon Thin Films. *J. Vac. Sci. Technol. A* **1997**, *15*, 936–940.

74. Lee, S.-H.; Lee, C.-S.; Lee, S.-C.; Lee, K.-H.; Lee, K.-R. Structural Properties of Amorphous Carbon Films by Molecular Dynamics Simulation. *Surf. Coat. Tech.* **2004**, *177 –178*, 812–817.

75. Belov, A. Yu.; Jager, H. U. Calculation of Intrinsic Stresses in Amorphous Carbon Films Grown by Molecular Dynamics Simulation: from Atomic to Macroscopic Scale. *Comp. Mater. Sci.* **2002**, *24*, 154–158.

76. Lu, Q.; Marks, N.; Schatz, G. C.; Belytschko, T. Nanoscale Fracture of Tetrahedral Amorphous Carbon by Molecular Dynamics: Flaw Size Insensitivity. *Phys. Rev. B* **2008**, *77*, 014109.

77. Lau, D. W. M.; McCulloch, D. G.; Marks, N. A.; Madsen, N. R.; Rode, A. V. High-Temperature Formation of Concentric Fullerene-Like Structures within Foam-Like Carbon: Experiment and Molecular Dynamics Simulation. *Phys. Rev. B* **2007**, *75*, 233408.